\documentclass{article}
\usepackage{graphicx}
\usepackage{amsmath}
\usepackage{geometry}
\usepackage{cite}
\usepackage{booktabs} 

\title{Multiscale Markowitz}
\author{Raphael Douady, Revant Nayar}
\date{August 2024}

\begin{document}

\maketitle
\begin{abstract}
    Traditional Markowitz portfolio optimization constrains daily portfolio variance to a target value, optimising returns, Sharpe or variance within this constraint. However, this approach overlooks the relationship between variance at different time scales, typically described by $\sigma(\Delta t) \propto (\Delta t)^{H}$ where $H$ is the Hurst exponent, most of the time assumed to be \(\frac{1}{2}\). This paper introduces a multifrequency optimization framework that allows investors to specify target portfolio variance across a range of frequencies, characterized by a target Hurst exponent $H_{target}$, or optimize the portfolio at multiple time scales. By incorporating this scaling behavior, we enable a more nuanced and comprehensive risk management strategy that aligns with investor preferences at various time scales. This approach effectively manages portfolio risk across multiple frequencies and adapts to different market conditions, providing a robust tool for dynamic asset allocation. This overcomes some of the traditional limitations of Markowitz, when it comes to dealing with crashes, regime changes, volatility clustering or multifractality in markets. We illustrate this concept with a toy example and discuss the practical implementation for assets with varying scaling behaviors.
\end{abstract}
\section{Introduction}

In the classical Markowitz portfolio optimization framework, the daily variance is typically constrained to a predefined maximum value, such that \(\sum_{i}\sigma_{i}^{2} < \sigma_{\text{target}}^{2}\). The objective is then to maximize returns while adhering to this variance constraint. This approach assumes a fixed variance at a single time scale, usually on a daily basis.

However, it is well understood that the variance of asset returns across different time scales is, to a large extent, related by a scaling law of the form:
\[
\sigma(\Delta t) \propto (\Delta t)^{H}
\]
where \(H = 0.5\) for Brownian motion \cite{mandelbrot1968fractional}. Despite this, the target portfolio variance is typically defined at only one frequency (commonly daily), without consideration for how variance behaves across other time scales.
In practice, it would be reasonable to expect that investors might have target portfolio variances across a range of frequencies, from the fast trade adjustment frequency to long-term portfolio objectives. The most general form of this would be an arbitrary function of scale, \(\sigma_{\text{target}}^{2}(\Delta t)\), which need not be continuous or differentiable. This function allows investors to specify their risk preferences over a spectrum of relevant time scales, for example, from one day to one month or more \cite{muzy2000modelling}.

For simplicity, however, we can assume self-similarity in the variance structure and instead fix the target scaling behavior, characterized by a target Hurst exponent \(H_{\text{target}}\). This gives us the relationship:
\[
\sigma_{\text{target}}^{2}(\Delta t) \propto (\Delta t)^{H_{\text{target}}}
\]
This approach allows for flexibility in portfolio optimization based on different investor risk preferences across time scales \cite{peters1994fractal}. For example, an investor who is more risk-averse to lower frequency volatility (e.g., weekly returns) than higher frequency volatility (e.g., daily returns) might prefer a target Hurst exponent \(H_{\text{target}} < H_{\text{market}}\). This situation could arise if a fund seeks to limit drawdowns at lower frequencies, where volatility is naturally higher. Conversely, if an investor is willing to tolerate more variance at lower frequencies, the target Hurst exponent might be set such that \(H_{\text{target}} > H_{\text{market}}\) \cite{gatheral2018volatility}.

In this article, we first justify the need for a multifrequency approach to time series analysis through observed stylized facts, such as varying volatility, fat tails and autocorrelation. Then we study a toy example based on multifractality and show how optimal weights in this context depend on the various parameters.

Finally, we evidence on US sector index tracking ETFs the superiority of multifrequency optimization over traditional Markowitz based on daily variances and covariances.

\subsection{Why Multiscale Optimization?}

We might question why there is a need at all to do multiscale portfolio optimization. Isn't the single scale case enough?

In reality, two facts occur in stock and stock index prices. On the one hand, non-trivial self-similarity with varying critical exponents has been observed in pricing time series since the time of Mandelbrot (who noticed it for cotton prices). At lower frequencies (weekly and below), momentum appears, i.e. a Hurst exponent above \(\frac{1}{2}\) while fat tails fade away. On the contrary, intraday returns display fatter tails but negative autocorrelation, i.e. a Hurst exponent below \(\frac{1}{2}\).

On the other hand, periods of high and low volatility operate like an acceleration and slowdown of the time clock. Using a fixed time clock for the return series boils down to using some kind of random time sampling with respect to the market "volatility time". We will se that, in fact, volatility is a hidden market variable. In other words, short of considering it as a hidden variable, prices series are not Markovian, despite the Efficient Market Hypothesis, which one to needs to enter into high frequency trading to invalidate it.

The logic of all this is that investors at marginally different time horizons behave in a similar way. Dynamic phase transitions are also characterised by self similar fixed points (markets show these near crashes and regime shifts). Motivated by this, there are certain cases where Markowitz breaks, particularly when the volatility spikes. Exposures that are calibrated on a low volatility period turn out be way to large for the newly appearing high volatility. We will show that this can be addressed by the multiscale optimization.

\subsection{Stylised Facts}

The following are typical stylised facts observed in the stock market and in financial markets in general:

\begin{itemize}
    \item \textbf{Volatility Clusters} Market volatility varies through time in a irregular and time-asymmetric manner. Volatility may suddenly surge following an event or a news that triggers its instability. Then it will take time to progressively decrease. 
    
    \item \textbf{Rough volatility:} This causes the volatility of some assets to increase disproportionately at longer scales, causing over-allocation to them \cite{gatheral2018volatility}. Globally, volatility remains in a limited range, implying a negative autocorrelation of its variations at various time scales.

    \item \textbf{Non-Ellipticity}  Asset returns exhibit varying fat-tailed distributions subject to the \emph{tail concentration effect}: under extreme conditions, the number of effective variables driving the market shrinks down to just a few of them. The Hurst exponent also varies, due to market impact eg. illiquids, corporate bonds, PE/VC funds, emerging markets etc. We know that Markowitz generally  overallocates to illiquids\cite{mandelbrot1963variation, cont2001empirical, amihud1986asset}.

    \item \textbf{Crashes and Bubbles:} During times of crashes, often Markowitz worsens drawdowns as opposed to an equally weighted portfolio \cite{cont2000herd}. This is because the volatility begins to scale non-trivially with scale. We expect that we see power laws during bubbles as well, with imaginary critical exponents.

    \item \textbf{Stochastic Volatility:} This introduces uncertainty on the volatility of the different assets, resulting in instabilities and estimation errors, in addition to non-generizability out of sample \cite{heston1993closed}.

    \item \textbf{Regime Changes:} It is unable to handle stochastic regime changes such as shocks induced by changes in interest rates, unemployment numbers, and global macroeconomic and political factors \cite{ang2012regime}. 

\end{itemize}

A lot of these are instances when the volatility becomes increasingly rough, with Hursts substantially different from half, or when tails become disproportionately fat \cite{gatheral2018volatility}. In the case of market crashes, in fact, that is what really happens. Bubbles too are characterized by Hursts, but those which are imaginary, corresponding to dynamics governed by the Log Periodic Power Law (LPPL). Markowitz is based on the condition of ellipticity, where all stocks have the same Hurst exponents. Therefore, optimal weights at one time scale generalize across scales.




\section{Motivations}
\subsection{Fractional Diffusion PDE}
We are motivated by the fact that pricing time series has been shown to exhibit self-similarity in various studies, starting with the time of Mandelbrot \cite{mandelbrot1963variation}. Also self-similar dynamics is at the endpoints of renormalization group flows. We start with the most general non-interacting PDE we can write down describing fractional diffusion in both space and time:
$\\\frac{\partial^{\beta}P(x,t)}{\partial t^{\beta}}=-K_{\alpha} (-\Delta)^{\alpha}P(x,t)\\$
Where
\begin{itemize}
    \item $P(x,t)$ is the probability density function

    \item $\frac{\partial^{\beta}}{\partial t^{\beta}}$ is fractional Caputo derivative (with $0<\beta<=1$)

    \item $(-\Delta)^{\alpha}$ is the fractional Laplacian (Riesz derivative) of order $\alpha$ (with $0<\alpha<=2$)

    \item $K_{\alpha}$ is a generalised diffusion coefficient. 
\end{itemize}
Here we see that there are qualitatively three types of cases:

\begin{itemize}
    \item Brownian motion ($\beta=1, \alpha=2$): This case equates to the usual Markowitz case. Here $|x| \propto t^{0.5}$

    \item Fat Tails ($\beta=1, 1<=\alpha<2$): This corresponds to no time dependence but fat tails with a power law $|x| \propto t^{1/\alpha}$

    \item Gaussian distribution with time dependence ($0<\beta<2, \alpha=2$): Here we have the benefit of no fat tails but time dependence causes anomalous scaling $|x| \propto t^{\beta}$

    \item Time dependence and Fat Tails($0<\beta<2, 1<=\alpha<=2$): This is the most general base with time dependence and fat tails. Here we have that $|x| \propto t^{\beta/\alpha}$ 
    
\end{itemize}
Note that in most studies, the Hurst is considered to be solely due to time dependence, and $H:=\beta/2$. However, this obfuscates the fact that anomalous scaling can be due to both time depedence and fat tails. Hence thereafter in this study, when we refer to the Hurst, we refer to the standardised Hurst defined as
$\\H :=\beta/\alpha\\$
This recognises that anomalous scaling can be due to both fat tails and time dependence. In a similar vein, in the presence of nonlinearities we can in principle have multifractality due to both the effect of the probability distribution ($\beta_{n}\neq n \beta$) and the interactions ($\alpha_{n} \neq \alpha/n$). Analogously, we have a standardised generalized Hurst:   
$\\H_{n}:=\beta_{n}/\alpha_{n}\\$
The direct observable here  is the scaling law $|x|^{n} \propto t^{H_{n}}$, while it is more subtle to address the contributions of the iid distribution and dependence structure to it, and will not be addressed. 

The discerning reader might be concerned here about the inconsistency of the Markowitz paradigm with the presence of fat tails, since the covariance matrix does not in principle converge under the presence of fat tails. The solution to that is straightforward: one can replace the covariance matrix with the $L^1$-modified covariance matrix and solve the optimisation problem:
$\\min_{w} w_{i}\Sigma^{L^1}_{ij} w_{j}\\$
Where the $L^1$-modified covariance matrix is written as:
$\\\Sigma^{L^1}_{ij}=\rho_{ij} |R_{i}-\tilde{R_{i}}||R_{j}-\tilde{R_{j}}|\\$
Where $\rho_{ij}$ are just the cross-asset correlations and $|R_{i}-\tilde{R_{i}}|$ and $|R_{j}-\tilde{R_{j}}|$ are absolute deviations from the robust central estimate (median). This will in principle have better convergence properties in the presence of tails, but should not make a substantial difference to the allocations. 
\section{Implementation}

In practice, it would work in the following manner. The returns are invariant under rescaling, so the return maximization condition will remain as is across a range of scales. As we progressively move to higher and higher frequencies, we will measure the variance of each asset across a range of frequencies, $\sigma_{i}^{2}(\Delta t)$, and then for each frequency we get:
\[
\sigma_{\text{portfolio}}(\Delta t)^{2} = \sum_{i} w_{i} w_{j} \Sigma_{ij}(\Delta t)
\]
We have a few options on how to implement this:
\begin{enumerate}
    \item For each frequency, we impose maximal return under the constraint \\ $\sigma_{\text{portfolio}}(\Delta t) < \sigma_{\text{target}}(\Delta t)$ where $\sigma^{2}_{\text{target}}(\Delta t) \propto \sigma^{2}_{\text{target}}(\Delta 1) |\Delta t|^{H_{target}}$. This approach ensures that risk, in the form of variance, is effectively managed across a range of frequencies \cite{berman2008multiscale}.
\item Construct a minimum variance or maximal Sharpe portfolios with the variance estimated by taking an average across scales or by averaging weights after multiscale optimization \cite{bianchi2015multiscale}. The alternative is to compute weights across various scales $w_{i}(\Lambda)$ and then average them out $\tilde{w} = \langle w_{i}(\Lambda) \rangle_{\Lambda}$ before renormalization.
\item Construct a minimum variance or maximal Sharpe portfolio where variance is a multiscale estimate of variance, involving the average over scales $\Sigma_{ij} = \langle \Sigma_{ij}(\Delta t)/(\Delta t) \rangle_{\Delta t}$.  

\end{enumerate}
We consider only the last case since the first does not sufficiently constrain the portfolio. We use the minimum variance optimisation, the advantage of which is that it directly measures the efficacy of the variance estimator without dependence on the mean estimator \cite{bianchi2015multiscale}.Let us see how multi-frequency optimization constrains our weights at a range of scales, beginning with special cases.

\section{Special Cases}
\subsection{Elliptical Case}

Consider a simplified scenario where all assets exhibit the same standardised Hurst exponent, \(H\), so $|x|=|t|^{H}$. Here, although we have anomalous scaling, the optimization problem can be solved at any given frequency, and the resulting solution will be applicable across all frequencies. This is due to the fact that when the scale is varied through a scaling transformation, the portfolio variance is modified by a factor of \((\Delta t_{1}/\Delta t_{2})^{H}\) \cite{mandelbrot1968fractional}, thus rescaling our optimisation function by a constant.


However, in practice, assets often exhibit diverse scaling behaviors with frequency and are typically not self-similar. In such cases, it is necessary to compute the optimization independently at each frequency to account for these differences \cite{muzy2000modelling}.
\subsection{General Case}

In a more generalized scenario, the variances and covariances of different assets, denoted as \(\sigma_{i}(\Delta t)\) and \(\rho_{i}(\Delta t)\), respectively, can vary arbitrarily with scale and are derived from empirical data. Similarly, the target variance \(\sigma_{\text{target}}(\Delta t)\) can be expressed as an arbitrary function of \(\Delta t\). Under these circumstances, the optimization problem is formulated as follows:

\begin{itemize}
    \item Minimize the portfolio variance: $\sum_{i} w_{i} \Sigma^{MS}_{ij} w_{j}$
    
    \item Here  $\Sigma^{MS}_{ij}=<\Sigma_{ij}(\Delta t)/\Delta t>$
    
    \item Ensure portfolio weights across scales sum to one: \(\sum_{i} w_{i}(\Lambda) = 1\)
    \item Maintain non-negative weights: \(w_{i}(\Lambda) \geq 0 \, \forall \, i, \Lambda \)
\end{itemize}

This general case highlights the complexity of portfolio optimization when dealing with assets that exhibit non-uniform scaling behavior across different time scales.

\section{Effect of Multifractality}

\subsection{Introduction to Multifractality}

Multifractality in financial time series reflects the idea that different parts of the data may exhibit different scaling behaviors. Unlike a monofractal process, which is characterized by a single Hurst exponent \(H\), a multifractal process is described by a spectrum of exponents \(H(q)\), where \(q\) is a moment order. This spectrum captures the complex, heterogeneous nature of financial markets, where the roughness of returns can vary depending on the time scale and the statistical moment being considered.

In a multifractal framework, the variance scaling law is generalized to:

\[
<\sigma^{q}(\Delta t)> \propto (\Delta t)^{H(q)},
\]

where \(H(q)\) is the multifractal scaling function, which depends on the moment \(q\). For example, \(H(1)\) corresponds to the traditional Hurst exponent used in variance scaling. Multifractality corresponds to $H(q)\neq q H(1)$

\subsection{Multifractal Portfolio Variance}

To incorporate multifractality into portfolio optimization, the variance of each asset \(i\) at a given time scale \(\Delta t\) is modeled using the multifractal formalism:

\[
\sigma_i^{q}(\Delta t) \propto \left(\Delta t\right)^{H_i(q)},
\]

where \(H_i(q)\) is the multifractal scaling exponent for asset \(i\) at moment \(q\). For Markowitz, we will be interested in $q=2$ where multifractality implies that $H(2) \neq 2 H(1)$. 
Similarly the covariance between stocks $i$ and $j$ scales as:

\[
\Sigma_{ij}(\Delta t) = \rho_{ij}(\Delta t) \sigma_i(\Delta t) \sigma_j(\Delta t)=\left(\Delta t\right)^{H_{ij}(2)},
\]
Where multifractality ensures $H_{ij}(2) \neq (H_{i}(1)+H_{j}(1))/2$. In fact from the above expression, if we assume a the scaling exponent for the correlation:
\[\rho_{ij}\propto (\Delta t)^{H^{\rho}_{ij}}\]
We find that:
\[H^{\rho}_{ij}=H_{ij}(2)-(H_{i}(1)+H_{j}(1))\]
This is well known to be non-zero and positive as a result of the Epps effect \cite{epps1979comovements}, where correlation between any two pairs of stocks is expected to increase with increasing length scale. Empirically from studies of the Epps effect, we find that $H^{\rho}_{ij}\approx 0.3$, which implies that the variation of cross correlation with scale decreases as one moves to lower and lower frequencies.  
The portfolio variance at time scale \(\Delta t\) then becomes:

\[
\sigma_{\text{portfolio}}^2(\Delta t) = \sum_{i=1}^{N} w_i^2 \sigma_i^2(\Delta t) + 2 \sum_{i=1}^{N} \sum_{j=i+1}^{N} w_i w_j \rho_{ij}(\Delta t) \sigma_i(\Delta t) \sigma_j(\Delta t).
\]

This expression now accounts for multifractality by incorporating the \(q\)-dependent scaling behavior of the assets, in terms of both the variance and covariance. 

\subsection{Multifractal Optimization Problem}

The optimization problem in a multifractal setting aims to minimize the portfolio variance across both time scales and moments. The optimization problem can be formulated as:

\[
\text{Minimize } \sum_{k=1}^K \sigma_{\text{portfolio}}^2(\Delta t_k, q),
\]
Here:
\[
      \sigma_{\text{portfolio}}^2(\Delta t, q) = \sum_{i} w_i^2 \sigma_i^2(\Delta t) + 2 \sum_{ij} w_i w_j \rho_{ij}(\Delta t) \sigma_i(\Delta t) \sigma_j(\Delta t)
\]
We here will expand the variances and covariances with the correct scaling properties. This is subject to:

\[
\sum_{i=1}^{N} w_i \mu_i \geq \mu_{\text{target}},
\]

\[
\sum_{i=1}^{N} w_i = 1,
\]

\[
w_i \geq 0, \quad \forall i.
\]

Here, the optimization takes into account not only the scaling behavior across different time scales \(\Delta t_k\) but also the varying roughness of the time series as captured by the multifractal spectrum \(H_i(q)\). This leads to a more complex, yet more accurate, description of portfolio risk that better reflects the true nature of financial markets.

\subsection{Estimation of Multifractal Parameters}

To implement the multifractal optimization framework, it is necessary to estimate the multifractal scaling exponents \(H_i(q)\) for each asset. This can be done using methods such as the multifractal detrended fluctuation analysis (MF-DFA) or the wavelet transform modulus maxima (WTMM) method. These techniques allow for the extraction of the multifractal spectrum from historical return data.

The estimated multifractal spectrum can then be used to calculate the portfolio variance at different time scales and moments, which serves as the input for the optimization problem.





\section{Sensitivity Analysis}
We can compute the sensitivity of the weights to the Hursts, multifractal Hursts etc. First let us evaluate the dependence on volatility. 

To find the optimal weights \( \mathbf{w} \) that minimize \( \sigma_p^2 \) subject to the constraint \( \mathbf{w}^\top \mathbf{1} = 1 \), we employ the method of Lagrange multipliers. The Lagrangian \( \mathcal{L} \) is given by:

\[
\mathcal{L}(\mathbf{w}, \lambda) = \mathbf{w}^\top \Sigma \mathbf{w} - \lambda (\mathbf{w}^\top \mathbf{1} - 1)
\]

where \( \lambda \) is the Lagrange multiplier associated with the budget constraint. We implement first order conditions. Take the derivative of \( \mathcal{L} \) with respect to \( \mathbf{w} \) and set it to zero:

\[
\frac{\partial \mathcal{L}}{\partial \mathbf{w}} = 2\Sigma \mathbf{w} - \lambda \mathbf{1} = 0
\]

Solving for \( \mathbf{w} \):

\[
2\Sigma \mathbf{w} = \lambda \mathbf{1} \quad \Rightarrow \quad \mathbf{w} = \frac{\lambda}{2} \Sigma^{-1} \mathbf{1}
\]
Using the budget constraint \( \mathbf{w}^\top \mathbf{1} = 1 \):

\[
\mathbf{w}^\top \mathbf{1} = \left( \frac{\lambda}{2} \Sigma^{-1} \mathbf{1} \right)^\top \mathbf{1} = \frac{\lambda}{2} \mathbf{1}^\top \Sigma^{-1} \mathbf{1} = 1
\]

Let \( S = \mathbf{1}^\top \Sigma^{-1} \mathbf{1} \), a scalar. Then:

\[
\frac{\lambda}{2} S = 1 \quad \Rightarrow \quad \lambda = \frac{2}{S}
\]

Substituting back into the expression for \( \mathbf{w} \):

\[
\mathbf{w} = \frac{1}{S} \Sigma^{-1} \mathbf{1}
\]

Thus, the optimal weights are:

\[
\mathbf{w} = \frac{\Sigma^{-1} \mathbf{1}}{\mathbf{1}^\top \Sigma^{-1} \mathbf{1}}
\]

\subsection{Effect of Variance}
To demonstrate that \textbf{increasing the variance} \( \sigma_k^2 \) of a specific asset \( k \) leads to a \textbf{decrease in its optimal weight} \( w_k \), while keeping all other parameters constant.

We want to now express the weight of asset $k$. From the optimal weights expression:

\[
w_k = \frac{(\Sigma^{-1} \mathbf{1})_k}{S} = \frac{s_k}{S}
\]

where \( s_k = (\Sigma^{-1} \mathbf{1})_k \) and \( S = \mathbf{1}^\top \Sigma^{-1} \mathbf{1} \).

Now we want to analyze the effect of increasing \( \sigma_k^2 \). We aim to compute the derivative \( \frac{\partial w_k}{\partial \sigma_k^2} \) and show that it is negative.

\[
\frac{\partial w_k}{\partial \sigma_k^2} = \frac{\partial}{\partial \sigma_k^2} \left( \frac{s_k}{S} \right) = \frac{\frac{\partial s_k}{\partial \sigma_k^2} S - s_k \frac{\partial S}{\partial \sigma_k^2}}{S^2}
\]

Computing \( \frac{\partial s_k}{\partial \sigma_k^2} \) and \( \frac{\partial S}{\partial \sigma_k^2} \)

Since \( \mathbf{s} = \Sigma^{-1} \mathbf{1} \), we have:

\[
\frac{\partial \mathbf{s}}{\partial \sigma_k^2} = -\Sigma^{-1} \left( \frac{\partial \Sigma}{\partial \sigma_k^2} \right) \Sigma^{-1} \mathbf{1}
\]

Given that \( \Sigma \) only depends on \( \sigma_k^2 \) through the \( (k,k) \)-element:

\[
\frac{\partial \Sigma}{\partial \sigma_k^2} = \mathbf{e}_k \mathbf{e}_k^\top
\]

where \( \mathbf{e}_k \) is the \( k \)-th standard basis vector. Thus:

\[
\frac{\partial \mathbf{s}}{\partial \sigma_k^2} = -\Sigma^{-1} \mathbf{e}_k \mathbf{e}_k^\top \Sigma^{-1} \mathbf{1}
\]

Specifically, the derivative of \( s_k \) is:

\[
\frac{\partial s_k}{\partial \sigma_k^2} = - (\Sigma^{-1})_{k k} s_k
\]

Similarly, the derivative of \( S \) is:

\[
\frac{\partial S}{\partial \sigma_k^2} = - s_k^2
\]

Lets get the final expression for the derivative. Substituting back:

\[
\frac{\partial w_k}{\partial \sigma_k^2} = \frac{- (\Sigma^{-1})_{k k} s_k S + s_k^3}{S^2} = \frac{s_k (- (\Sigma^{-1})_{k k} S + s_k^2)}{S^2}
\]

Given that \( \Sigma^{-1} \) is positive definite, \( (\Sigma^{-1})_{k k} > 0 \), and \( S > s_k^2 \), it follows that:

\[
\frac{\partial w_k}{\partial \sigma_k^2} < 0
\]
We know that $\frac{\partial \sigma}{\partial H_{i}}=|\Delta t|^{H_{i}} ln(\Delta t)$

It follows that
\[
\frac{\partial w_k}{\partial H_{k}} < 0
\]
\subsection{Effect of Correlation and Multifractality}

Next, we consider the effect of increasing the correlation between two assets, say \( i \) and \( j \), on their combined weight \( w_i + w_j \). The covariance between assets \( i \) and \( j \) is given by:
\[
\sigma_{ij} = \rho_{ij} \sigma_i \sigma_j
\]
where \( \rho_{ij} \) is the correlation between assets \( i \) and \( j \). Increasing \( \rho_{ij} \) increases the covariance \( \sigma_{ij} \).

The weights \( w_i \) and \( w_j \) depend on the inverse of the covariance matrix \( \Sigma^{-1} \). To compute the effect of increasing \( \rho_{ij} \), we differentiate the weights  $w_i$ and $w_j$ with respect to \( \rho_{ij} \):
\[
\frac{\partial (w_i)}{\partial \rho_{ij}} = \frac{\partial (s_i)}{\partial \rho_{ij}} \cdot \frac{1}{S} - (s_i) \cdot \frac{\partial S}{\partial \rho_{ij}} \cdot \frac{1}{S^2}\]
\[
\frac{\partial (w_j)}{\partial \rho_{ij}} = \frac{\partial (s_j)}{\partial \rho_{ij}} \cdot \frac{1}{S} - (s_j) \cdot \frac{\partial S}{\partial \rho_{ij}} \cdot \frac{1}{S^2}
\]
where \( s_i = (\Sigma^{-1} \mathbf{1})_i \) and \( s_j = (\Sigma^{-1} \mathbf{1})_j \). The derivative \( \frac{\partial \Sigma}{\partial \rho_{ij}} \) has nonzero entries only at positions \( (i,j) \) and \( (j,i) \).
Here we can have that $w_{i}$ and $w_{j}$ might increase or decrease. However, if we do an optimisation over the combined asset $w_{i}+w_{j}$ we can conclude that the combined weight of the synthetic asset decreases with an increase in correlation. This is because in this modified picture, all you are doing is changing the variance of the two assets:
$\\<lr_{synthetic,ij}^{2}>=<(lr_{i}+lr_{j})^{2}>=<(lr_{i})^{2}>+<(lr_{j})^{2}>+<lr_{i} lr_{j}>\\$
So all the other variances and covariances are the same, except that of the synthetic combination of $i$ and $j$. Thus if we freeze $w_{ij}$, we get from the previous result:
\[\frac{\partial w_{ij}}{\partial \rho_{ij}}<=0
\]
Thus:
\[\frac{\partial w_{ij}}{\partial H_{ij}}=\frac{\partial w_{ij}}{\partial \rho_{ij}} \frac{\partial \rho_{ij}}{\partial H_{ij}}=\frac{\partial w_{ij}}{\partial \rho_{ij}}*|\Delta t|^{H_{ij}} ln(\Delta t)<0
\]
Here note that although empirically $H_{ij}>0$ which causes a positive multifractality, one could have had $H_{ij}<0$ as well in principle. In that case the correlations would decay with scale, and hence the weight of the synthetic asset would increase with an increase in the multifractal parameter $H_{ij}$. 
\subsection{Summary}
To summarize our results we get as expected:
\begin{itemize}
\item If you increase volatility of a stock, its portfolio weight goes down
    \item If you increase Hurst ($\beta/2$) of a stock,$H_{i}$ goes up, its portfolio weight goes down
    \item If you increase the fat tailedness of a stock ($1/\alpha$),$H_{i}$ goes up, its portfolio weight goes down
    \item If you increase correlation between two stocks, their combined weight goes down.
    \item If you introduce positive multifractality in two stocks' correlation (increased Epps effect),$H_{ij}$ goes up, their combined weight goes down; and vice versa.  
\end{itemize}
\section{Out of Sample Performance}

To rigorously assess the efficacy of the proposed multiscale optimization methods, we conducted an out-of-sample evaluation in two scenarios: the minimum variance portfolio and the maximal Sharpe portfolio, with the mean estimated using a simple moving average. The portfolio weights correspond to allocations across the 11 sectors of the S\&P 500. The following choices were used for the backtest:
\begin{itemize}
    \item We utilized five years of data, spanning from 2019 to 2024, a period that notably includes significant market events such as the March 2020 COVID-19 crash and the 2022 market correction.
    
    \item The SPDR sector ETFs were used in lieu of allocation to the 11 sectors of the $S\&P 500$, making it a long-only sector rotation strategy. We also considered a factor rotation strategy taking 9 factors (quality, growth, value, low volatility etc) where each factor was represented by ETFs.

    \item A lookback of six months was chosen (125 days), which is common in such studies.

    \item Minimum variance portfolios with overlapping vs non-overlapping averages were considered at lower frequencies

    \item At each scale, the covariance matrix was evaluated on the same period, but lower frequency covariances were averaged over all posible non-overlapping sets to increase robustness. We compute covariance matrix at a range of scales $\Sigma_{ij}^{MS}=<\Sigma_{ij}(\Delta t)/\Delta t>_{\Delta t}$ 
    \item Transaction costs were assumed to be negligible for the purpose of this study.

\end{itemize}
\begin{table}[]
  \centering
  \caption{Backtest Results for Various Portfolio Optimization Methods (Sector Rotation)}
    \begin{tabular}{lccc}
      \hline
      Method & Sharpe Ratio & Sortino Ratio & Max Drawdown (\%) \\
      \hline
      Equally Weighted & 0.45 & 0.53 & -39.9 \\
      Traditional Markowitz & 0.35 & 0.41 & -33.1 \\
      Multiscale Markowitz & 0.53 & 0.62 & -31.5 \\
      Multiscale Markowitz (Overlapping) & 0.49 & 0.58 & -30.3 \\
      \hline
    \end{tabular}%
\end{table}

\begin{table}[]
  \centering
  \caption{Backtest Results for Various Portfolio Optimization Methods (Factor Rotation)}
    \begin{tabular}{lccc}
      \hline
      Method & Sharpe Ratio & Sortino Ratio & Max Drawdown (\%) \\
      \hline
      Equally Weighted & 0.42 & 0.52 & -39.1 \\
      Traditional Markowitz & 0.43 & 0.49 & -36.9 \\
      Multiscale Markowitz & 0.53 & 0.61 & -36.3 \\
      Multiscale Markowitz (Overlapping) & 0.57 & 0.66 & -34.9 \\
      \hline
    \end{tabular}%
  
\end{table}
Our findings indicate that the multiscale optimization method results in portfolios with higher Sharpe ratios and Sortino ratios, as well as lower kurtosis and drawdowns. This enhanced performance is attributable to the multiscale method's ability to account for the effects of non-ellipticity and fat tails, phenomena that are often inadequately captured by traditional minimum variance portfolios \cite{calvet2002multifractality, bianchi2015multiscale}.

\end{document}